\documentstyle[sprocl,cite]{article}

\input{psfig}
\def\be{\begin{equation}}
\def\ee{\end{equation}}
\def\bea{\begin{eqnarray}}
\def\eea{\end{eqnarray}}

\begin{document}

\title{ORIGIN AND PROPAGATION OF THE HIGHEST ENERGY COSMIC RAYS}

\author{ R.J. PROTHEROE }

\address{Department of Physics and Mathematical Physics\\
The University of Adelaide, Adelaide, Australia 5005}

\maketitle\abstracts{
\begin{center}
ABSTRACT\\
\end{center}
In this lecture I give an overview of shock acceleration,
interactions of high energy cosmic rays with, and propagation 
through, the background radiation, and the resulting electron-photon cascade.
I argue that while the origin of the highest energy cosmic rays
is still uncertain, it is not necessary to invoke exotic models
such as emission by topological defects to explain the existing
data.
It seems likely that shock acceleration at Fanaroff-Riley Class II
radio galaxies can account for the existing data.
However, new cosmic ray data, as well as better estimates of the 
extragalactic radiation fields and magnetic fields will be necessary
before we will be certain of the origin of the highest energy particles
occurring in nature.
}

\section{Introduction}

Cosmic rays with energies
up to 100 TeV are thought to arise predominantly through 
shock acceleration by supernova remnants (SNR) in our Galaxy\cite{Lag83}.
% this 
% hypothesis has recently gained support from direct evidence for the 
% first time\cite{Stu95,Osb95}. 
A fraction of the cosmic rays accelerated 
 should interact within the supernova remnant and 
produce  gamma--rays \cite{DruryAharonianVolk,GaisserProtheroeStanev96}, and 
 recent observations above 100 MeV by the EGRET instrument on the Compton 
 Gamma Ray Observatory have found gamma ray signals 
 associated with at least two supernova remnants -- IC~443 and
 $\gamma$~Cygni \cite{Esposito96}
(however, it is possible that the gamma ray emission from IC~443 is
associated with a pulsar within the remnant rather than the remnant 
itself \cite{Brazier96}). 
 Further evidence for acceleration in SNR
 comes from the recent ASCA observation of non-thermal X--ray emission
 from SN~1006\cite{Koyama95}.
 Reynolds\cite{Reynolds96} and Mastichiadis\cite{Mastichiadis96}
 interpret the latter as synchrotron emission by 
 electrons accelerated in the remnant up to energies as high as 100 TeV,
although Donau and Biermann\cite{DonauBiermann96} suggest it may be bremsstrahlung
from much lower energy electrons.

Acceleration to somewhat higher energies than 100 TeV may be possible\cite{Mar90}, 
but probably not high enough to explain the smooth extension of the spectrum
to 1 EeV. 
Several explanations for the
origin of the cosmic rays in this energy range have been suggested: 
reacceleration of the supernova component while still
inside the remnant\cite{Axf91}; by several supernovae exploding
into a region evacuated by a pre-supernova star\cite{Ip91};  or acceleration in 
shocks inside the strong winds from hot
stars or groups of hot stars\cite{Bie93}.
At 5 EeV the spectral slope changes, and there is evidence
for a lightening in composition\cite{Bir94} and it is likely 
this marks a change from galactic cosmic rays to  
extragalactic cosmic rays being dominant.

The cosmic ray air shower events with the highest energies so far detected
have energies of $2 \times 10^{11}$ GeV\cite{Hay94} and 
$3 \times 10^{11}$ GeV\cite{Bir95}.
The question of the origin of these cosmic rays having energy significantly
above $10^{11}$ GeV is complicated by
propagation of such energetic particles through the universe.
Nucleons interact with the cosmic background radiation fields, 
losing energy by pion photoproduction, and may emerge as either protons
or neutrons with reduced energy.
The threshold for pion photoproduction on the microwave background  is 
$\sim 2 \times 10^{10}$ GeV, and at $3 \times 10^{11}$ GeV 
the energy-loss distance is about 20 Mpc.
Propagation of cosmic rays over substantially larger distances
gives rise to a cut-off in the spectrum at $\sim 10^{11}$ GeV 
as was first shown by Greisen\cite{Gre66}, and 
Zatsepin and Kuz'min\cite{Zat66}, the ``GZK cut-off'', and 
a corresponding pile-up at slightly lower energy\cite{Hil85,Ber88}.
These processes occur not only during propagation, but also during acceleration
and may actually limit the maximum energies particles can achieve.

In this lecture I give an overview of shock acceleration,
describe interactions of high energy protons and nuclei with radiation,
discuss maximum energies obtainable during acceleration,
outline propagation of cosmic rays through the background radiation
and the consequent electron-photon cascading, 
and finally discuss conventional and exotic models of the highest energy 
cosmic rays.

%%%%%%%%%%%%%%%%%%%%%%%%%%%%%%%%%%%%%%%%%%%%%%%%%%%%%%%%%%%%%%%%%%%%%%%%%%%%%%%

\section{Cosmic Ray Acceleration}
%%%%%%%%%%%%%%%%%%%%%%%%%%%%%%%%%%%%%%%%%%%%%%%%%%%%%%%%%%%%%%%%%%%%%%%%%%%%%%%
%%%%%%%%%%%%%%%%%%%%%%%%%%%%%%%%%%%%%%%%%%%%%%%%%%%%%%%%%%%%%%%%%%%%%%%%%%%%%%%

For stochastic particle acceleration by electric fields induced by
motion of magnetic fields $B$, the rate of energy gain by relativistic
particles of charge $Ze$ can be written (in SI units)
\begin{equation}
\left. {dE \over dt} \right|_{\rm acc} = \xi Ze c^2 B
\end{equation}
where $\xi < 1$ and depends on the acceleration mechanism.
I shall give a simple heuristic treatment of Fermi acceleration
based on that given in Gaisser's excellent book\cite{Gaisser}.
I shall start with 2nd order Fermi acceleration (Fermi's original theory) and
describe how this can be modified in the context of supernova shocks,
or other strong astrophysical shocks, into the more efficient
1st  order Fermi mechanism at supernova (SN) or other shocks.
More detailed and rigorous treatments are given in several
review articles\cite{Drury,BlandfordEichler87,BerezhkoKrymsky88,JonesEllison91}.
See the review by Jones and Ellison\cite{JonesEllison91} on the plasma
physics of shock acceleration which also includes a brief historical review
and refers to early work.

\subsection{Fermi's Original Theory}

Gas clouds in the interstellar medium have random velocities of $\sim 15$ km/s 
superimposed on their regular motion around the galaxy.
Cosmic rays gain energy on average when scattering off these magnetized clouds.
A cosmic ray enters a cloud and scatters off irregularities in the magnetic field 
which is tied to the  cloud because it is partly ionized.

In the frame of the cloud: (a) there is no change in energy because the
scattering is collisionless, and so there is elastic 
scattering between the cosmic ray 
and the cloud as a whole which is much more massive than the cosmic ray; 
(b) the cosmic ray's direction
is randomized by the scattering and it emerges from the cloud in a random direction.

\begin{figure}[htb]
%\rule{5cm}{0.2mm}\hfill\rule{5cm}{0.2mm}
%\vskip 2.5cm
%\rule{5cm}{0.2mm}\hfill\rule{5cm}{0.2mm}
\psfig{figure=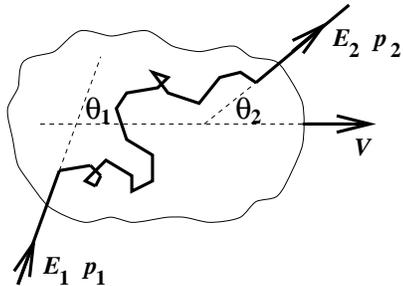,height=1.5in}
\caption{Interaction of cosmic ray of energy $E_1$ with ``cloud'' moving with
speed $V$
\label{fig:fermi_acc_orig}}
\end{figure}

Consider a cosmic ray entering a cloud with energy $E_{1}$ and 
momentum $p_{1}$ travelling in a
direction making angle $\theta_{1}$ with the cloud's direction.
After scattering inside the 
cloud, it emerges with energy $E_{2}$ and momentum $p_{2}$
at angle $\theta_{2}$ to the cloud's direction (Fig.~\ref{fig:fermi_acc_orig}).
The energy change is obtained by 
applying the Lorentz transformations 
between the laboratory frame (unprimed)
and the cloud frame (primed).
Transforming to the cloud frame:
\begin{equation}
E_{1}^{\prime} = \gamma E_{1} (1 - \beta \cos \theta_{1})
\end{equation}
where $\beta = V/c$ and $\gamma = 1/\sqrt{1-\beta^{2}}$.

\noindent Transforming to the laboratory frame:
\begin{equation}
E_{2} = \gamma E_{2}^{\prime} (1 + \beta \cos \theta_{2}^{\prime}).
\end{equation}
Since $E_{2}^{\prime} =  E_{1}^{\prime}$ we obtain the fractional change in energy
$(E_{2}-E_{1})/E_{1}$,
\begin{equation}
{\Delta E \over E} = 
{1 - \beta \cos \theta_{1} + \beta \cos \theta_{2}^{\prime}
- \beta^{2} \cos \theta_{1} \cos \theta_{2}^{\prime} \over
1 - \beta^{2}} -1.
\end{equation}

We need to obtain average values of $ \cos \theta_{1}$ and $ \cos \theta_{2}^{\prime}$.
Inside the cloud, the cosmic ray scatters off magnetic irregularities many times 
so that its direction is randomized,
\begin{equation}
\langle \cos \theta_{2}^{\prime} \rangle =0.
\end{equation}
The average value of cos$\theta_{1}$ depends on the rate at which cosmic rays collide with clouds at different angles.
The rate of collision is proportional to the relative velocity
between the cloud and the particle so that
the probability per unit solid angle of having a collision at angle
$\theta_{1}$ is proportional to $(v - V \cos \theta_{1})$.  
Hence, for ultrarelativistic particles ($v=c$)
\begin{equation}
{dP \over d \Omega_{1}} \propto (1 - \beta \cos \theta_{1}),
\end{equation}
and we obtain
\begin{equation}
\langle \cos \theta_{1} \rangle = 
\int \cos \theta_{1} {dP \over d \Omega_{1}} d \Omega_{1} /
\int {dP \over d \Omega_{1}} d \Omega_{1} = - {\beta \over 3},
\end{equation} 
giving
\begin{equation}
{\langle \Delta E \rangle \over E} = 
{1 + \beta^{2}/3 \over
1 - \beta^{2}} -1 \simeq {4 \over 3} \beta^{2}
\end{equation}
since $\beta \ll 1$.

We see that $\langle \Delta E \rangle / E \propto \beta^{2}$ 
is positive (energy gain), but is 2nd order in $\beta$ and
because $\beta \ll 1$ the average energy gain is very small.  
This is because there are
almost as many overtaking collisions (energy lost) as there are head-on collisions
(energy gain).

%%%%%%%%%%%%%%%%%%%%%%%%%%%%%%%%%%%%%%%%%%%%%%%%%%%%%%%%%%%%%%%%%%%%%%%%%%%%

\subsection{1st Order Fermi Acceleration at SN or Other Shocks}

Fermi's original theory was modified in the 
1970's\cite{Axford77,Krymsky77,Bell78,BlandfordOstriker78} 
to describe more efficient acceleration
(1st order in $\beta$) taking place at supernova shocks
but is generally applicable to strong shocks in other astrophysical contexts.
 
During a supernova explosion several solar masses of material are ejected at 
a speed of $\sim 10^{4}$ km/s which is much faster than the speed of sound in the
interstellar medium (ISM) which is $\sim$ 10 km/s.
A strong shock wave propagates radially out as the ISM 
and its associated magnetic field piles up in front of the
supernova ejecta.
The velocity of the shock, $V_{S}$, depends on the velocity of the ejecta, 
$V_{P}$, and on the
ratio of specific heats, $\gamma$.
The SN will have ionized the surrounding gas which will therefore be
monatomic ($\gamma = 5/3$), and theory of shock hydrodynamics
shows that for a strong shock
\begin{equation}
V_{S}/V_{P} \simeq 4/3.
\end{equation}

\begin{figure}[htb]
%\rule{5cm}{0.2mm}\hfill\rule{5cm}{0.2mm}
%\vskip .5cm
%\rule{5cm}{0.2mm}\hfill\rule{5cm}{0.2mm}
\psfig{figure=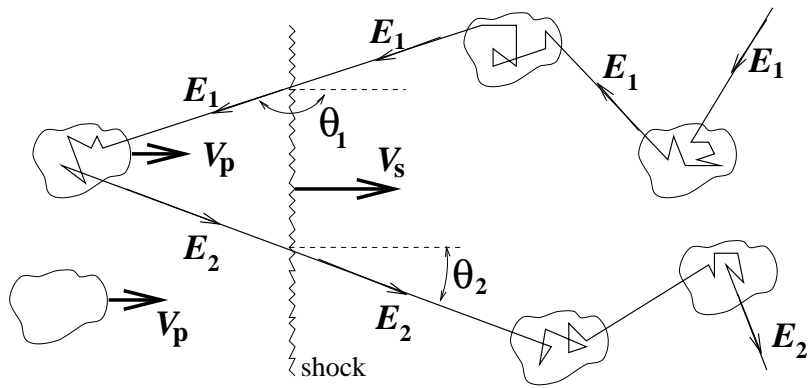,height=2.0in}
\caption{Interaction of cosmic ray of energy $E_1$ with a shock moving with 
speed $V_s$.
\label{fig:fermi_acc_shock}}
\end{figure}

In order to work out the energy gain per shock crossing, 
we can visualize magnetic irregularities on either side of the shock as clouds
of magnetized plasma of Fermi's original theory (Fig.~\ref{fig:fermi_acc_shock}).
By considering the rate at which cosmic rays cross the shock from downstream to
upstream, and upstream to downstream, one finds
$\langle \cos \theta_{1} \rangle = -2/3$ and
$\langle \cos \theta_{2}^{\prime} \rangle = 2/3$,
giving
\begin{equation}
{\langle \Delta E \rangle \over E}  \simeq {4 \over 3} \beta \simeq {V_{S} \over c}.
\end{equation}
Note this is 1st order in $\beta$ and is therefore more efficient than 
Fermi's original theory.
This is because of the converging flow -- whichever side of the shock you are
on, if you are moving with the plasma, the plasma on the other side of the shock
is approaching you at speed $V_p$.

\begin{figure}[htb]
%\rule{5cm}{0.2mm}\hfill\rule{5cm}{0.2mm}
%\vskip .5cm
%\rule{5cm}{0.2mm}\hfill\rule{5cm}{0.2mm}
\psfig{figure=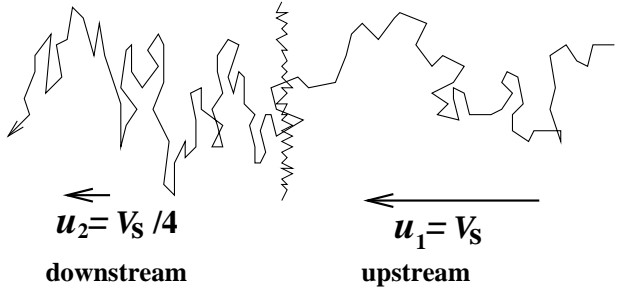,height=1.5in}
\caption{Diffusion of cosmic rays from upstream to downstream seen in the 
shock frame
speed $V_s$.
\label{fig:up_downstream}}
\end{figure}

To obtain the energy spectrum we need to find the probability of 
a cosmic ray encountering
the shock once, twice, three times, etc.
If we look at the diffusion of a cosmic ray as seen in the rest frame of the shock
(Fig.~\ref{fig:up_downstream}),
there is clearly a net flow of the energetic particle population 
in the downstream direction.
The net flow rate gives the rate at which cosmic rays are lost downstream
\begin{equation}
R_{{\rm loss}} = n_{\rm CR} V_{S}/4 \hspace{10mm} \rm m^{{-2}} s^{{-1}}
\end{equation}
since cosmic rays with number density $n_{\rm CR}$ at the shock
are advected downstream with speed $V_{S}/4$ (from right to left in 
Fig.~\ref{fig:up_downstream}).

Upstream of the shock, cosmic rays travelling at speed $v$ at angle $\theta$ to 
the shock normal (as seen in the laboratory frame) approach the shock with
speed $(V_{S} +  v \cos \theta)$ as seen in the shock frame. 
Clearly, to cross
the shock, $\cos \theta > -V_{S}/v$.
Then, assuming cosmic rays upstream are isotropic, the rate at which they cross
from upstream to downstream is 
\begin{equation}
R_{\rm cross} =  n_{\rm CR} {1 \over 4 \pi} \int_{-V_S/v}^1 
(V_S +  v \cos \theta) 2 \pi d( \cos \theta) =  n_{\rm CR} v/4 \hspace{5mm} \rm m^{-2} s^{-1}.
\end{equation}

The probability of crossing the shock once and then escaping from the shock 
(being lost downstream) is the ratio of these two rates:
\begin{equation}
{\rm Prob.(escape)} = R_{\rm loss}/R_{\rm cross} = V_{S}/v
\end{equation}
where we have neglected relativistic transformations of the rates because $V_S \ll c$.
The probability of returning to the shock after crossing from upstream to downstream
is
\begin{equation}
{\rm Prob.(return)} = 1 - {\rm Prob.(escape)}
\end{equation}
and so the probability of returning to the shock $k$ times and also of
crossing the shock at least $k$ times is
\begin{equation}
{\rm Prob.(cross} \ge k{\rm )} = [1 - {\rm Prob.(escape)}]^{k}.
\end{equation}
Hence, the energy after $k$ shock crossings is
\begin{equation}
E = E_{0} \left( 1 + {\Delta E \over E} \right)^{k}
\end{equation}
where $E_{0}$ is the initial energy.

To derive the spectrum, we note that 
the integral energy spectrum (number of particles with energy greater than $E$)
on acceleration must be
\begin{equation}
Q(>E) \propto  [1 - {\rm Prob.(escape)}]^{k}
\end{equation}
where 
\begin{equation}
k = {\ln (E/E_{0}) \over \ln (1 + \Delta E/E)}.
\end{equation}
Hence,
\begin{equation}
\ln Q(>E) = A + {\ln (E/E_{0}) \over \ln (1 + \Delta E/E)} 
\ln [1 - {\rm Prob.(escape)}],
\end{equation}
where $A$ is a constant, and so
\begin{equation}
\ln Q(>E) = B - \Gamma \ln E
\end{equation}
where $B$ is a constant and
\begin{equation}
\Gamma = - {\ln [1 - {\rm Prob.(escape)}] \over \ln (1 + \Delta E/E)} \approx 1.
\end{equation}

Hence we arrive at the spectrum of cosmic rays on acceleration
\begin{equation}
Q(>E) \propto E^{{-1}} \hspace{1cm} \rm (integral \; form)
\end{equation}
\begin{equation}
Q(E) \propto E^{{-2}} \hspace{1cm} \rm (differential \; form).
\end{equation}
The observed cosmic ray spectrum is steepened by energy-dependent escape
of cosmic rays from the Galaxy.

\subsection{Shock Acceleration Rate}

The rate of gain of energy is given by 
\begin{equation}
\left. {dE \over dt} \right|_{\rm acc} = {\Delta E \over t_{\rm cycle}}
\end{equation}
where $t_{\rm cycle}$ is the time for one complete cycle, i.e. 
from crossing the shock from upstream to downstream, diffusing back towards the 
shock and crossing from downstream to upstream, and finally returning to the shock.
We shall discuss this process in the shock frame (see Fig.~\ref{fig:up_downstream})
and consider first particles crossing the shock from upstream to downstream and 
diffusing back to the shock, i.e. we shall work out the average time spent
downstream.
Since we are considering non-relativistic shocks, the time scales are 
approximately the same in the upstream and downstream plasma frames, and so 
in this section I shall drop the use of subscripts indicating the frame
of reference.

Diffusion takes place in the presence of advection at speed $u_2$ in
the downstream direction.
The typical distance a particle diffuses in time t is $\sqrt{k_2t}$ where
$k_2$ is the diffusion coefficient in the downstream region.
The distance advected in this time is simply $u_2t$.
If $\sqrt{k_2t} \gg u_2t$ the particle has a very high probability of returning to 
the shock, and if $\sqrt{k_2t} \ll u_2t$ the particle has a very high 
probability of 
never returning to the shock (i.e. it has effectively escaped downstream).
So, we set $\sqrt{k_2t} = u_2t$ to define a distance $k_2/u_2$ downstream
of the shock which is effectively a boundary between the region closer to
the shock where the particles will usually return to the shock and the region
farther from the shock in which the particles will usually be advected downstream
never to return.
There are $n_{\rm CR} k_2/u_2$ particles per unit area of shock between
the shock and this boundary.
Dividing this by $R_{\rm cross}$ we obtain the average time spent downstream
before returning to the shock
\begin{equation}
t_2 \approx {4 \over c} {k_2 \over u_2}.
\end{equation}

Consider next the other half of the cycle after the particle has crossed 
the shock from downstream to upstream until it returns to the shock.
In this case we can define a boundary at a distance $k_1/u_1$ upstream
of the shock such that nearly all particles upstream of this boundary 
have never encountered the shock, and nearly all the particles between
this boundary and the shock have diffused there from the shock.
Then dividing the number of particles per unit area of shock between
the shock and this boundary, $n_{\rm CR} k_1/u_1$, by  $R_{\rm cross}$ 
we obtain the average time spent upstream before returning to the shock
\begin{equation}
t_1 \approx {4 \over c} {k_1 \over u_1},
\end{equation}
and hence the cycle time
\begin{equation}
t_{\rm cycle} \approx {4 \over c} \left( {k_1 \over u_1} + {k_2 \over u_2} \right).
\end{equation}
The acceleration time at energy $E$, defined by $E/(dE/dt)$ is then
given by
\begin{equation}
t_{\rm acc} \approx {4 \over u_1} \left( {k_1 \over u_1} + {k_2 \over u_2} \right).
\end{equation}

We next consider the diffusion for the cases of parallel, oblique, and 
perpendicular shocks, and estimate the maximum acceleration rate for these cases.
The diffusion coefficients required $k_1$ and $k_2$ are the coefficients for
diffusion parallel to the shock normal.
The diffusion coefficient along the magnetic field direction is 
some factor $\eta$ times the minimum diffusion coefficient, known
as the Bohm diffusion coefficient,
\begin{equation}
k_\parallel = \eta {1 \over 3} r_g c
\end{equation}
where $r_g$ is the gyroradius, and $\eta > 1$.

Parallel shocks are defined such that the shock normal is parallel to
the magnetic field ($\vec{B} || \vec{u_1}$).
In this case, making the approximation that $k_1 = k_2 = k_\parallel$ 
and $B_1 = B_2$ one obtains 
\begin{equation}
t_{\rm acc}^\parallel \approx {20 \over 3} {\eta E \over e B_1 u_1^2}.
\end{equation}
For a shock speed of $u_1 = 0.1 c$ and $\eta=10$ one obtains an acceleration
rate (in SI units) of 
\begin{equation}
\left. {dE \over dt} \right|_{\rm acc} \approx 1.5 \times 10^{-4} e c^2 B.
\end{equation}

For the oblique case, the angle between the magnetic field direction 
and the shock normal is different in the upstream and downstream regions,
and the direction of the plasma flow also changes at the shock.
The diffusion coefficient in the direction at angle $\theta$ 
to the magnetic field direction is given by
\begin{equation}
k = k_\parallel \cos^2 \theta + k_\perp \sin^2 \theta
\end{equation}
where $k_\perp$ is the diffusion coefficient perpendicular to the magnetic field.
Jokipii\cite{Jokipii87} shows that 
\begin{equation}
k_\perp \approx {k_\parallel \over 1 + \eta^2}
\end{equation}
provided that $\eta$ is not too large (values in the range up to 10 appear
appropriate).

In the case of acceleration at perpendicular shocks, Jokipii\cite{Jokipii87}
has shown that acceleration can be much faster than for the parallel case.
For $k_{xx} = k_\perp$ and $B_2 \approx 4 B_1$ and one obtains
\begin{equation}
t_{\rm acc}^\perp \approx {8 \over 3} {E \over \eta e B_1 u_1^2}.
\end{equation}
For a shock speed of $u_1 = 0.1 c$ and $\eta=10$ one obtains an acceleration
rate (in SI units) of 
\begin{equation}
\left. {dE \over dt} \right|_{\rm acc} \approx 0.04 e c^2 B.
\end{equation}

This discussion of shock acceleration has been of necessity brief,
and has omitted a number of subtleties such as the finite thickness
of the shock front, and the reader is referred to the excellent reviews
cited earlier for such details.
Nevertheless, the basic concepts have been described in sufficient detail
that we can consider acceleration and interactions of the highest energy 
cosmic rays, and to what energies they can be accelerated.
Supernova shocks remain strong enough to continue accelerating cosmic rays for 
about 1000 years.  The rate at which cosmic rays are accelerated is inversely
proportional to the diffusion coefficient (faster diffusion means less time near
the shock).  For the maximum feasible acceleration rate, a typical interstellar 
magnetic field, and 1000 years for acceleration, 
energies of $10^{{14}} \times Z$ eV are possible ($Z$ is atomic number)
at parallel shocks and $10^{{16}} \times Z$ eV at perpendicular shocks.

%%%%%%%%%%%%%%%%%%%%%%%%%%%%%%%%%%%%%%%%%%%%%%%%%%%%%%%%%%%%%%%%%%%%%%%%%%%%%%%
%%%%%%%%%%%%%%%%%%%%%%%%%%%%%%%%%%%%%%%%%%%%%%%%%%%%%%%%%%%%%%%%%%%%%%%%%%%%%%%

\section{Interactions of High Energy Cosmic Rays}

Interactions of cosmic rays with radiation and magnetic fields are important
both during acceleration when the resulting energy losses compete with
energy gains by, for example, shock acceleration, and during propagation 
from the acceleration region to the observer.
For ultra-high energy (UHE) cosmic rays, the most important processes are 
pion photoproduction  and Bethe-Heitler pair production both
on the microwave background, and synchrotron radiation.
In the case of nuclei, photodisintegration on the microwave background
is important.
In this section I shall concentrate on photoproduction  and pair production.

The mean interaction length, $x_{p \gamma}$, of a proton of energy $E$ is given by,
\begin{equation}
        [x_{p \gamma}(E)]^{-1}= {1 \over 8 \beta E^2}
\int_{\varepsilon_{\rm min}}^
        {\infty} \frac{n(\varepsilon)}
        {\varepsilon^2} \int_{s_{\rm min}}^{s_{\rm max}(\varepsilon,E)} 
        \sigma(s)(s-m_p^2 c^4)ds d\varepsilon,
        \label{eq:mpl}
\end{equation}
where $n(\varepsilon)$ is the differential photon number density
of photons of energy $\varepsilon$, and 
$\sigma(s)$ is the appropriate total cross section for the process 
in question for a centre of momentum (CM) frame energy squared, $s$, given by
\begin{equation}
s=m_p^2 c^4 + 2 \varepsilon E(1 - \beta \cos \theta)
\label{eq:s}
\end{equation}
where $\theta$ is the angle between the directions of the 
proton and photon,
and $\beta c$ is the proton's velocity.

For pion photoproduction  
\begin{equation}
s_{\rm min} = (m_pc^2+m_{\pi}c^2)^2 \approx 1.16 \; \rm GeV^2,
\end{equation}
and
\begin{equation}
\varepsilon_{\rm min}= {m_{\pi}c^2(m_{\pi}c^2+2m_pc^2) \over 2E(1+\beta)}
\approx {m_{\pi}c^2(m_{\pi}c^2+2m_pc^2) \over 4E}.
\end{equation}
For photon-proton pair-production the threshold is somewhat lower,
\begin{equation}
s_{\rm min}= (m_pc^2 + 2 m_ec^2)^2 \approx0.882 \; \rm GeV^2,
\end{equation}
and
\begin{equation}
\varepsilon_{\rm min}\approx m_ec^2(m_ec^2+m_pc^2)/E.
\end{equation}
For both processes,
\begin{equation}
s_{\rm max}(\varepsilon,E) = m_p^2c^4+2\varepsilon E(1+\beta)
\approx m_p^2c^4+4\varepsilon E,
\end{equation} 
and $s_{\rm max}(\varepsilon,E)$ corresponds to a head-on collision
of a proton of energy $E$ and a photon of energy $\varepsilon$.

Examination of the integrand in Equation \ref{eq:mpl} shows that the 
energy of the soft photon interacting with a proton 
of energy $E$ is distributed as
\begin{equation}
p(\varepsilon) = \frac{x_{p \gamma}(E)n(\varepsilon)}{8 \beta E^2 \varepsilon^2}
        \Phi(s_{\rm max}(\varepsilon,E))
\end{equation}
in the range $\varepsilon_{\rm min} \leq \varepsilon \leq \infty$ where 
\begin{equation}
        \Phi(s_{\rm max})=\int_{s_{\rm min}}^{s_{\rm max}}\sigma(s)
        (s-m_p^2c^4)ds.
\end{equation}
Similarly, examination of the integrand in Equation \ref{eq:mpl} shows that the
square of the
total CM frame energy is distributed as
\begin{equation}
        p(s) =\frac{\sigma(s)(s-m_p^2c^4)}{\Phi(s_{\rm max})},
\end{equation}
in the range $s_{\rm min}\leq s\leq s_{\rm max}$.

The Monte Carlo rejection technique can be used to 
sample $\varepsilon$ and $s$ respectively
from the two distributions, and Equation \ref{eq:s} is used to find $\theta$.
One then Lorentz transforms the interacting particles to the frame in which
the interaction is treated (usually the proton rest frame),
and samples momenta of particles produced 
in the interaction from the appropriate differential cross section 
by the rejection method.
The energies of produced particles are then Lorentz transformed to the
laboratory frame, and 
the final energy of the proton is obtained by requiring energy conservation.
In this procedure, it is not always possible to achieve exact conservation
of both momentum and energy while sampling particles from inclusive
differential cross sections (e.g. multiple pion production well above threshold), 
and the momentum of the last particle sampled is 
therefore adjusted to minimize the error.

\begin{figure}[htb]
%\rule{5cm}{0.2mm}\hfill\rule{5cm}{0.2mm}
%\vskip 2.5cm
%\rule{5cm}{0.2mm}\hfill\rule{5cm}{0.2mm}
\psfig{figure=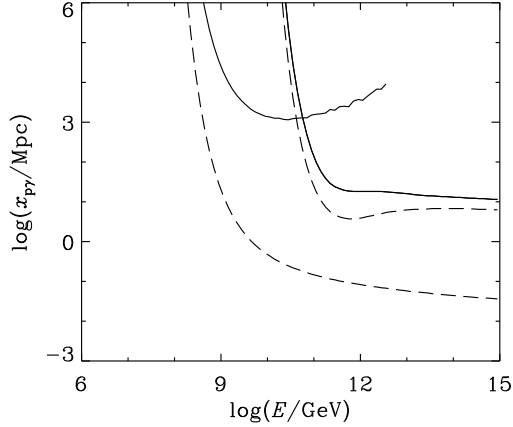,height=2.5in}
\caption{Mean interaction length (dashed lines) and energy-loss 
distance (solid lines), $E/(dE/dx)$, for proton-photon pair-production
and pion-production in the microwave background
(lower and higher energy curves respectively). 
(From Protheroe and Johnson\protect \cite{ProtheroeJohnson95}).
\label{fig:pgpiee3k_xloss}}
\end{figure}

The mean interaction lengths for both processes, $x_{p \gamma}(E)$,  are obtained 
from Equation \ref{eq:mpl} for interactions in the microwave background 
and are plotted as dashed lines in Fig. \ref{fig:pgpiee3k_xloss}.
Dividing by the inelasticity, $\kappa(E)$, one obtains the energy-loss 
distances for the two processes,
\begin{equation}
{E \over dE/dx} = {x_{p \gamma}(E) \over \kappa(E)}.
\end{equation}
% Energy-loss distances from the work of Berezinsky and Grigor'eva\cite{Ber88}, 
% Yoshida and Teshima\cite{Yos93} and Rachen and Biermann\cite{Rac93},
% and Protheroe and Johnson\cite{ProtheroeJohnson95} are compared
% in Fig. \ref{fig:xloss_comp}

\section{Maximum Energies}

Protons and nuclei can be accelerated to
much higher energies than electrons for a given magnetic environment.
For stochastic particle acceleration by electric fields induced by
motion of magnetic fields $B$, the rate of energy gain by relativistic
particles of charge $Ze$ can be written (in SI units)
\begin{equation}
\left. {dE \over dt} \right|_{\rm acc} = \xi Ze c^2 B
\label{eq:max_acc}
\end{equation}
where $\xi < 1$ and depends on the acceleration mechanism;  a value
of $\xi =0.04$ might be achieved by first order Fermi acceleration
at a perpendicular shock with shock speed $\sim 0.1 c$.  

\subsection{Limits from Synchrotron Losses}

The rate of energy loss by synchrotron radiation of a particle of mass
$Am_p$, charge $Ze$, and energy $\gamma m c^2$ is
\begin{equation}
- \left. {dE \over dt} \right|_{\rm syn} = {4 \over 3} \sigma_T
\left( {Z^2 m_e \over Am_p} \right)^2 {B^2 \over 2 \mu_0} \gamma^2 c.
\end{equation}
Equating the rate of energy gain with the rate of energy loss by
synchrotron radiation places one limit on the maximum energy achievable
by electrons, protons and nuclei:
\begin{eqnarray}
E_e^{\rm max} &=& 6.0 \times 10^{2} \xi^{1/2}
\left( {B \over {\rm 1 \; T}} \right)^{-1/2} \hspace{5mm} {\rm GeV}, \\
E_p^{\rm max} &=& 2.0 \times 10^{9} \xi^{1/2}
\left( {B \over {\rm 1 \; T}} \right)^{-1/2} \hspace{5mm} {\rm GeV}, \\
E_{Z,A}^{\rm max} &=& 2.0 \times 10^{9} \xi^{1/2} {A^2 \over Z^{3/2}}
\left( {B \over {\rm 1 \; T}} \right)^{-1/2} \hspace{5mm} {\rm GeV}.
\end{eqnarray}
Other limits on the maximum energy are placed by the dimensions of
the acceleration region and the time available for acceleration.  These 
limits were obtained and discussed in some detail by 
Biermann \& Strittmatter\cite{BS87}.

\subsection{Limits from Interactions with Radiation}

\begin{figure}[htb]
%\rule{5cm}{0.2mm}\hfill\rule{5cm}{0.2mm}
%\vskip 2.5cm
%\rule{5cm}{0.2mm}\hfill\rule{5cm}{0.2mm}
\psfig{figure=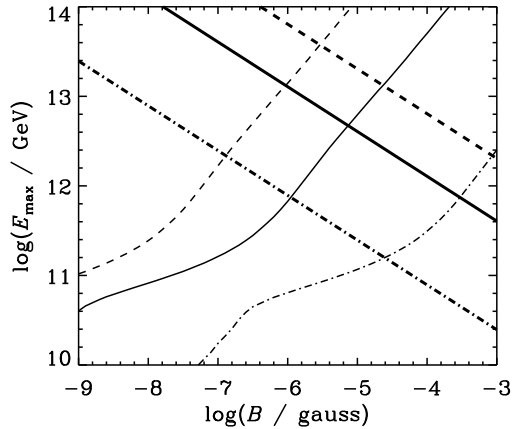,height=2.5in}
\caption{Maximum proton energy as a function of magnetic field (see text).
\label{fig:pgpiee3k_emax}}
\end{figure}

Equating the total energy loss rate for proton--photon interactions
(i.e. the sum of pion production and Bethe-Heitler pair production)
in Fig~\ref{fig:pgpiee3k_xloss} 
to the rate of energy gain by acceleration gives the maximum proton energy
in the absence of other loss processes.
This is shown in Fig~\ref{fig:pgpiee3k_emax} 
as a function of magnetic field which determines the 
rate of energy gain through Eq.~\ref{eq:max_acc}.
The result is shown by the thin curves for the maximum possible 
acceleration rate $\xi=1$ (dashed), plausible acceleration at perpendicular
shock $\xi=0.04$ (solid), and plausible acceleration at parallel shock
$\xi=1.5 \times 10^{-4}$ (dot-dash).
Also shown is the maximum energy determined by synchrotron losses (thick lines)
for the three cases.
As can be seen, for a perpendicular shock it is possible to accelerate protons
to $\sim 4 \times 10^{12}$ GeV in a $\sim 10^{-5}$ G field.

\begin{figure}[htb]
%\rule{5cm}{0.2mm}\hfill\rule{5cm}{0.2mm}
%\vskip 2.5cm
%\rule{5cm}{0.2mm}\hfill\rule{5cm}{0.2mm}
\psfig{figure=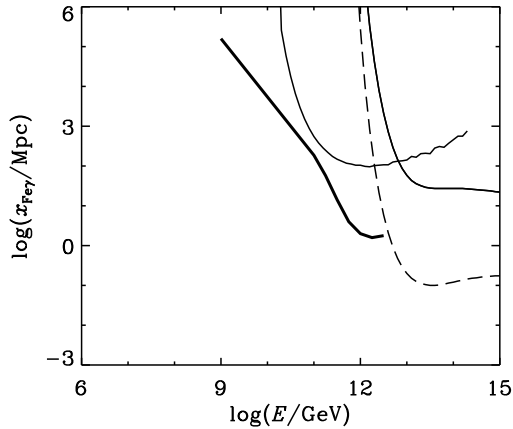,height=2.5in}
\caption{Mean interaction length (dashed line) and energy-loss 
distance (solid line), $E/(dE/dx)$, for Fe-photon pion-production, 
and energy-loss distance for pair-production in the microwave background
(leftmost solid line).
Also shown is the photodisintegration distance (thick solid curve).
\label{fig:fepiee_xloss}}
\end{figure}
\begin{figure}[htb]
%\rule{5cm}{0.2mm}\hfill\rule{5cm}{0.2mm}
%\vskip 2.5cm
%\rule{5cm}{0.2mm}\hfill\rule{5cm}{0.2mm}
\psfig{figure=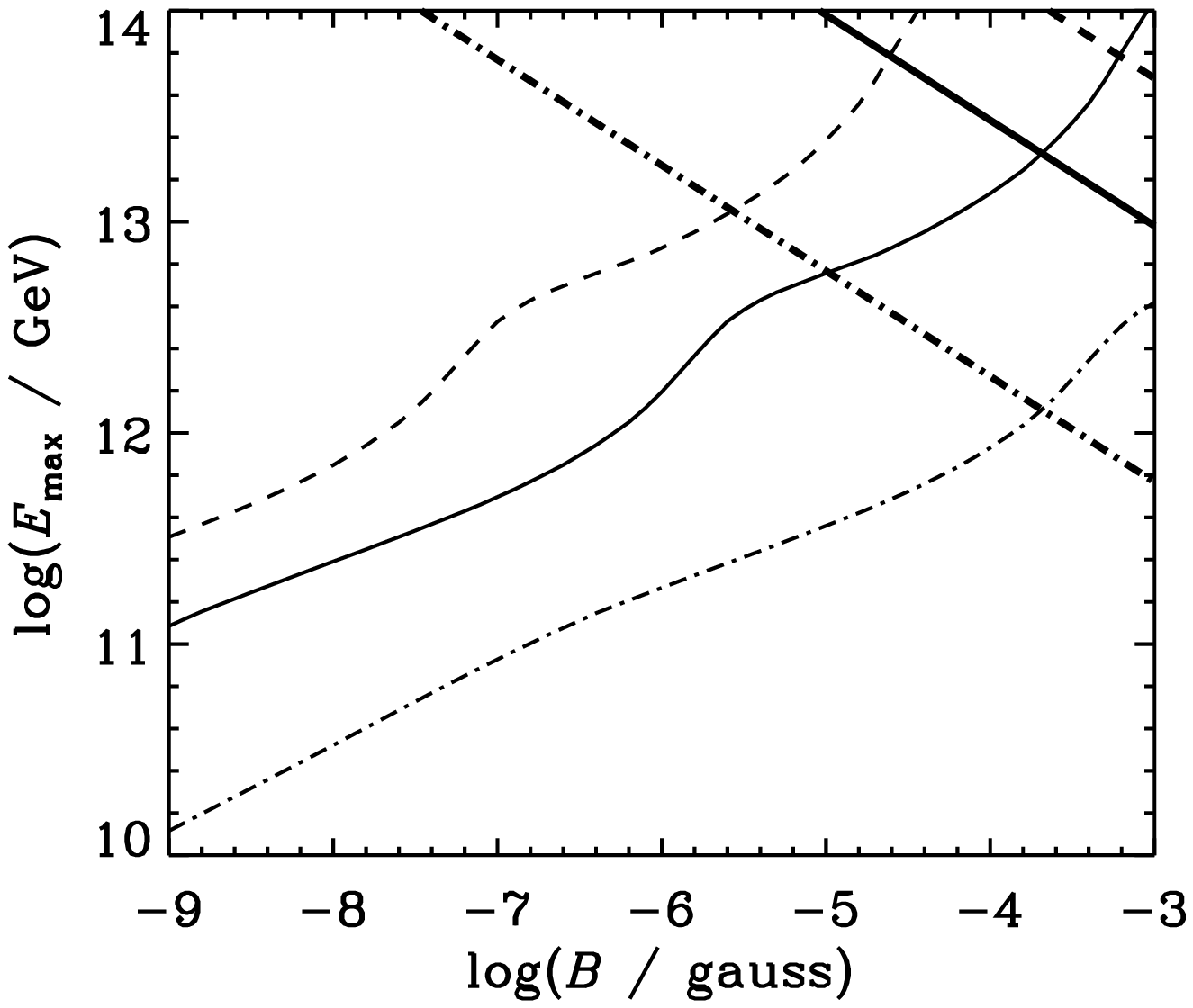,height=2.5in}
\caption{Maximum iron nucleus energy as a function of magnetic field
(see Fig. \protect \ref{fig:pgpiee3k_emax}).
\label{fig:fe_emax}}
\end{figure}

In the case of nuclei the situation is a little more complicated.
The threshold condition for Bethe-Heitler pair production can be expressed as
\begin{equation}
\gamma > {m_e c^2 \over \varepsilon} \left( 1 + {m_e \over Am_p} \right),
\end{equation}
and the  threshold condition for pion photoproduction can be expressed as
\begin{equation}
\gamma > {m_\pi c^2 \over 2 \varepsilon} \left( {1 + {m_\pi \over 2A m_p}} \right).
\end{equation}
Since $\gamma = E / A m_p c^2$, where $A$ is the mass number, we will need to
shift both energy-loss distance 
curves in Fig.~\ref{fig:pgpiee3k_xloss} to higher energies by a factor of $A$.
We shall also need to shift the curves up or down as discussed below.

For Bethe-Heitler pair production the energy lost by a nucleus in each collision
near threshold is approximately $\Delta E \approx \gamma 2 m_e c^2$. 
Hence the inelasticity is 
\begin{equation}
K \equiv { \Delta E \over E} \approx {2m_e \over A m_p},
\end{equation}
and is a factor of $A$ lower than for protons.
On the other hand, the cross section goes like $Z^2$, so the overall shift 
is down (to lower energy-loss distance) by $Z^2/A$.  
For example, for iron nuclei the energy loss distance for pair production 
is reduced by a factor $26^2/56 \approx 12.1$.

For pion production the energy lost by a nucleus in each collision
near threshold is approximately $\Delta E \approx \gamma m_\pi c^2$,
and so, as for pair production, the inelasticity is factor $A$ lower than 
for protons.
The cross section increases approximately as $A^{0.9}$ giving an overall
increase in the energy loss distance for pion production of a factor 
$\sim A^{0.1} \approx 1.5$ for iron nuclei.

The energy loss distances for pair production and pion photoproduction,
together with the mean free path  for pion photoproduction are shown for
iron nuclei in Fig.~\ref{fig:fepiee_xloss}.
Photodisintegration is very important and has been considered in detail by
Tkaczyk W. et al.\cite{Tka75} and Puget et al.\cite{Pug76}.
The photodisintegration distance defined by $A/(dA/dx)$ taken from 
Puget et al.\cite{Pug76} is also shown in Fig.~\ref{fig:fepiee_xloss}.
Since iron nuclei will be fragmented also during pion photoproduction, 
the effective loss distance will be given by the photodisintegration distance below
$\sim 3 \times 10^{12}$ GeV, and by the {\em mean free path}  
for pion photoproduction at higher energies.
The effective loss distance given in Fig.~\ref{fig:fepiee_xloss} is used
together with the acceleration rate and synchrotron
loss rate for iron nuclei to obtain the maximum energy as a function
of magnetic field.
This is shown in Fig.~\ref{fig:fe_emax} which is analogous to 
Fig.~\ref{fig:pgpiee3k_emax} for protons.
We see that for a perpendicular shock it is possible to accelerate 
iron nuclei to $\sim 2 \times 10^{13}$ GeV in a $\sim 2 \times 10^{-4}$ G
field.
While this is higher than for protons, iron nuclei are likely
to get photodisintegrated  into nucleons of maximum energy 400 EeV,
and so there is not much to be gained unless the source is nearby.
Of course, potential acceleration sites need to have the appropriate 
combination of size (much larger than the gyroradius at the maximum energy),
magnetic field, and shock velocity (or other relevant velocity), and
these criteria have been discussed in detail by Hillas\cite{Hil84}.

%%%%%%%%%%%%%%%%%%%%%%%%%%%%%%%%%%%%%%%%%%%%%%%%%%%%%%%%%%%%%%%%%%%%%%%%%%%%%%%
 
\section{Cascading During Propagation}

There are several cascade processes which are important for UHE
cosmic rays propagating over large distances through a radiation field:
protons interact with photons resulting in  
pion production and pair production; electrons interact via
inverse-Compton scattering and triplet pair production, 
and emit synchrotron radiation in the intergalactic magnetic field; 
$\gamma$-rays interact by pair production.
Energy losses due to cosmological redshifting 
of high energy particles and 
$\gamma$-rays can also be important, and the cosmological redshifting of the
background radiation fields means that energy thresholds and interaction
lengths for the above processes also change with epoch
(see e.g. Protheroe et al.\cite{ProtheroeStanevBerezinsky}).

\begin{figure}[htb]
%\rule{5cm}{0.2mm}\hfill\rule{5cm}{0.2mm}
%\vskip 2.5cm
%\rule{5cm}{0.2mm}\hfill\rule{5cm}{0.2mm}
\psfig{figure=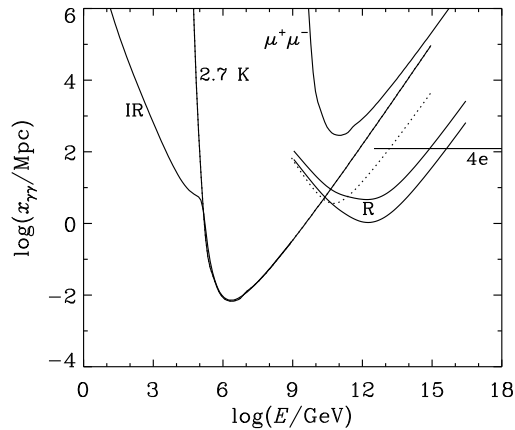,height=2.5in}
\caption{The mean interaction length for pair production for
$\gamma$-rays in the Radio Background calculated in the present work
(solid curves labelled R:  upper curve -- no evolution of normal galaxies;
lower curve -- pure luminosity evolution of normal galaxies)
and in the radio background of Clark \protect
\cite{Cla70} (dotted line).  Also shown are the mean interaction length
for pair production in the microwave background (2.7K), the infrared and
optical background (IR), and muon pair production ($\mu^+\mu^-$) and
double pair production (4e) in the microwave background
\protect\cite{ProtheroeJohnson95}.
(From Protheroe and Biermann\protect \cite{ProtheroeBiermann96a}).
\label{fig:ggee_radio}}
\end{figure}

The energy density of the extragalactic background radiation is dominated by 
that from the cosmic microwave background at a temperature of 2.735 K.
Other components of the extragalactic background radiation are discussed in 
the review of Ressel and Turner\cite{Res90}. 
The extragalactic radiation fields important for cascades initiated by UHE 
cosmic rays include the cosmic microwave background,
the  radio background and the infrared--optical background. 
The radio background
was measured over twenty years ago\cite{Cla70,Bri67}, but  
the fraction of this radio background which is truly extragalactic, 
and not contamination from our own Galaxy, is still debatable.
Berezinsky\cite{Ber69} was first to calculate the mean free path on the 
radio background.
Recently Protheroe and Biermann\cite{ProtheroeBiermann96a}
have made a new calculation of the extragalactic radio background
radiation down to kHz frequencies.
The main contribution to the background is from normal galaxies and is
uncertain due to uncertainties in their evolution.
The mean free path of photons in this radiation field as well as in the 
microwave and infrared backgrounds is shown in 
Fig.~\ref{fig:ggee_radio}.
Also shown is the mean interaction length for 
muon pair-production which is negligible in comparison with interactions 
with pair production on the radio background and double pair production
on the microwave background.

\begin{figure}[htb]
%\rule{5cm}{0.2mm}\hfill\rule{5cm}{0.2mm}
%\vskip 2.5cm
%\rule{5cm}{0.2mm}\hfill\rule{5cm}{0.2mm}
\psfig{figure=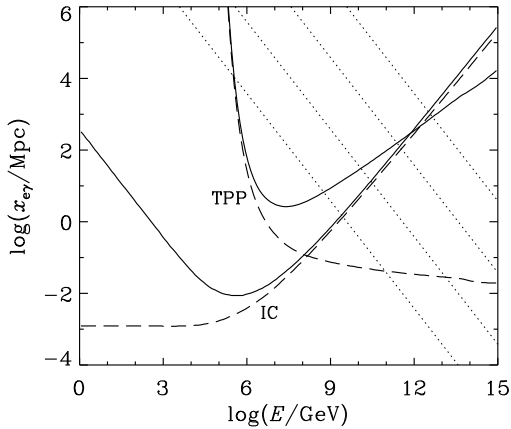,height=2.5in}
\caption{The mean interaction length (dashed line) and energy-loss distance 
(solid line), $E/(dE/dx)$, for electron-photon triplet
pair production (TPP) and inverse-Compton scattering (IC) in the 
microwave background. 
The energy-loss distance for synchrotron radiation is also
shown (dotted lines) for intergalactic magnetic fields of $10^{-9}$ (bottom),
$10^{-10}$, $10^{-11}$, and $10^{-12}$ gauss (top).
(From Protheroe and Johnson\protect \cite{ProtheroeJohnson95}).
\label{fig:e3kmfp}}
\end{figure}

Inverse Compton interactions of high energy electrons 
and triplet pair production 
can be modelled by the Monte Carlo technique\cite{Pro86,Pro90,Pro92,Mas94}, 
and the mean interaction lengths and energy-loss distances for these 
processes are given in Fig. \ref{fig:e3kmfp}.
Synchrotron losses must also be included in calculations and the
energy-loss distance has been added to Fig. \ref{fig:e3kmfp} for various
magnetic fields.

%%%%%%%%%%%%%%%%%%%%%%%%%%%%%%%%%%%%%%%%%%%%%%%%%%%%%%%%%%%%%%%%%%%%%%%%%%%%%%%
\subsection{Practical Aspects of the Cascade}

%%%%%%%%%%%%%%%%%%%%%%%%%%%%%%%%%%%%%%%%%%%%%%%%%%%%%%%%%%%%%%%%%%%%%%%%%%%%%%%
Where possible, to
take account of the exact energy dependences of cross-sections,
one can use the Monte Carlo method.
However, direct application of Monte Carlo techniques to cascades dominated
by the physical processes described above 
over cosmological distances takes excessive computing time. 
Another approach based on the matrix multiplication method has been described
by Protheroe\cite{Pro86} and developed in later 
papers \cite{Pro93,ProtheroeJohnson95}. 
A Monte Carlo program is used to calculate the yields of secondary particles
due to interactions with radiation, and 
spectra of produced pions are decayed (e.g. using routines in SIBYLL\cite{Fle94})
to give yields of gamma-rays, electrons and neutrinos.
The yields are then used to build up transfer matrices which
describe the 
change in the spectra of particles produced after propagating through 
the radiation fields for a distance $\delta x$.
Manipulation of the transfer matrices as described below enables one to 
calculate the spectra of particles resulting from propagation 
over arbitrarily large distances.

%%%%%%%%%%%%%%%%%%%%%%%%%%%%%%%%%%%%%%%%%%%%%%%%%%%%%%%%%%%%%%%%%%%%%%%%%%%%%%%
\subsection{Matrix Method}
%%%%%%%%%%%%%%%%%%%%%%%%%%%%%%%%%%%%%%%%%%%%%%%%%%%%%%%%%%%%%%%%%%%%%%%%%%%%%%%
In the work of 
Protheroe and Johnson\cite{ProtheroeJohnson95},
fixed logarithmic energy bins were used, 
and the energy spectra of particles of type $\alpha$
($\alpha = \gamma,e,p,n,\nu_e,\bar{\nu}_e,\nu_\mu,\bar{\nu}_\mu$)
at distance $x$ in the cascade
are represented by vectors $F_{j}^{\alpha}(x)$
which give the total number of particles of type $\alpha$
in the $j$th energy bin at distance $x$.
Transfer matrices, T$_{ij}^{\alpha\beta}(\delta x)$, give the
number of particles of type $\beta$ in the bin $j$
which result at a distance
$\delta x$ after a particle of type $\alpha$ and
energy in bin $i$ initiates a cascade.
Then, given the spectra of particles at distance $x$ one can obtain the spectra
at distance $(x + \delta x)$
\begin{equation}
F_{j}^{\beta}(x +\delta x) = \sum_{\alpha}
        \sum_{i=j}^{180} {\rm T}_{ij}^{\alpha \beta}
        (\delta x)F_{i}^{\alpha}(x)
\end{equation}
where $F_{i}^{\alpha}(x)$ are the input spectra
(number in the $i$th energy bin) of species $\alpha$.

We could also write this as
\begin{equation}
[{\rm F}(x +\delta x)] = [{\rm T}(\delta x)][{\rm F}(x)]
\end{equation}
where 
\begin{equation}
\rm
[F] = \left[ \begin{array}{c} F^{\gamma} \\ F^{e} \\ F^{p} 
\\ \vdots  \end{array} \right] , \hspace{5mm}
[T] = \left[ \begin{array}{cccc} 
{\rm T}^{\gamma \gamma} &  {\rm T}^{e \gamma} & 
         {\rm T}^{p \gamma} &  \cdots \\
{\rm T}^{\gamma e} &  {\rm T}^{ee} &  {\rm T}^{pe} & \cdots \\
{\rm T}^{\gamma p} &  {\rm T}^{ep} &  {\rm T}^{pp} & \cdots \\
\vdots  & \vdots  & \vdots & \ddots \end{array} \right]
. 
\end{equation}

The transfer matrices depend on particle yields, $\rm Y_{ij}^{\alpha\beta}$, 
which are  defined as the probability of
producing a particle of type $\beta$ 
in the energy bin $j$ when a primary particle of type $\alpha$ 
with energy in bin $i$ undergoes an interaction.
To calculate $\rm Y_{ij}^{\alpha\beta}$ a Monte Carlo simulation can be used
(see Protheroe and Johnson\cite{ProtheroeJohnson95} for details).

%%%%%%%%%%%%%%%%%%%%%%%%%%%%%%%%%%%%%%%%%%%%%%%%%%%%%%%%%%%%%%%%%%%%%%%%%
\subsection{Matrix Doubling}
%%%%%%%%%%%%%%%%%%%%%%%%%%%%%%%%%%%%%%%%%%%%%%%%%%%%%%%%%%%%%%%%%%%%%%%%%
From Fig.~\ref{fig:e3kmfp}
we see that the smallest effective interaction length
is that for synchrotron losses by electrons at high energies.
We require $\delta x$ be much smaller than this distance which is
of the order of parsecs for the highest magnetic field considered.
To follow the cascade for a distance corresponding to a redshift
of $z \sim 9$, and to complete the calculation of the cascade 
using repeated application of the transfer matrices 
would require $ \sim  10^{12}$ steps.
This is clearly impractical, and one must use the more sophisticated approach
described below.

The matrix method and matrix doubling technique have been used for many years 
in radiative transfer problems\cite{Hul63,Hov71}.
The method described by Protheroe and Stanev\cite{Pro93}
is summarized below. 
Once the transfer matrices have been calculated for a distance
$\delta x$, the transfer matrix for a distance $2\delta x$ is simply 
given by applying the transfer matrices twice, i.e.
\begin{equation}
[{\rm T}(2 \delta x)] = [{\rm T}(\delta x)]^2.
\end{equation}
In practice, it is necessary to use high-precision during computation
(e.g. double-precision in FORTRAN), and to ensure
that energy conservation is preserved after each doubling.
The new matrices may then be used to
calculate the transfer matrices for distance $4\delta x$, $8\delta x$, 
and so on.
A distance $2^n\delta x$ only requires the application of this
`matrix doubling' $n$ times. 
The spectrum of electrons and photons after a large
distance $\Delta x$ is then given 
by
\begin{equation}
[{\rm F}(x +\Delta x)] = [{\rm T}(\Delta x)][{\rm F}(x)]
\end{equation}
where $[{\rm F}(x)]$ represents the input spectra, and $\Delta x=2^n\delta x$.
In this way, cascades over long distances can be modelled quickly and 
efficiently.

%%%%%%%%%%%%%%%%%%%%%%%%%%%%%%%%%%%%%%%%%%%%%%%%%%%%%%%%%%%%%%%%%%%%%%%%%%%%%%%

\section{The Origin of Cosmic Rays between 100 TeV  and 300 EeV}
%%%%%%%%%%%%%%%%%%%%%%%%%%%%%%%%%%%%%%%%%%%%%%%%%%%%%%%%%%%%%%%%%%%%%%%%%%%%%%%

The subject of the origin of cosmic rays at these energies
has been reviewed\cite{Hil84,Axf94}, and one of the very few plausible 
acceleration sites may be associated with 
the radio lobes of powerful radio galaxies, either in the hot 
spots\cite{RachenBiermann93} or possibly the cocoon or jet\cite{Nor95}. 
One-shot processes comprise another possible class of sources
\cite{Has92,Sor87}. 

Acceleration at the termination shock of the galactic wind from our Galaxy
has been also been suggested\cite{Jok85}, 
but due to the lack of any statistically significant anisotropy 
associated with the Galaxy is unlikely to be the explanation. 
However, a very recent re-evaluation of the world
data set of cosmic rays has shown that there is a 
correlation of the arrival directions of cosmic rays
above 40 EeV with the supergalactic plane\cite{Sta95}, lending
support to an extragalactic origin above this energy, and in particular
to models where ``local'' sources ($<100$ Mpc) would appear to cluster
near the supergalactic plane (e.g. powerful radio galaxies 
as in the model of Rachen and Biermann\cite{RachenBiermann93}.

\begin{figure}[htb]
%\rule{5cm}{0.2mm}\hfill\rule{5cm}{0.2mm}
%\vskip 2.5cm
%\rule{5cm}{0.2mm}\hfill\rule{5cm}{0.2mm}
\psfig{figure=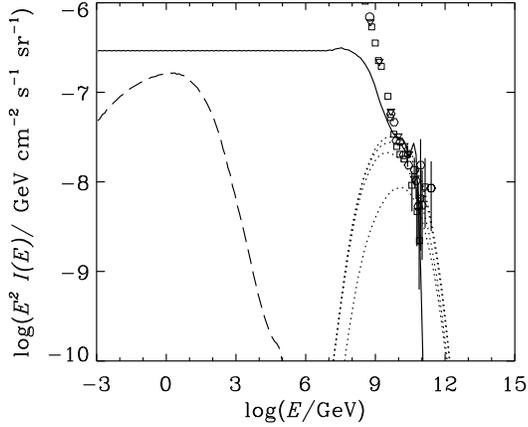,height=2.5in}
\caption{Cosmic ray proton intensity multiplied by $E^{2}$ 
in the model of Rachen and Biermann for $H_0=75$ km s$^{-1}$ Mpc$^{-1}$
with proton injection up to $3 \times 10^{11}$ GeV (solid line).  
Also shown are intensities of neutrinos (dotted lines,
$\nu_\mu , \bar{\nu}_\mu, \nu_e, \bar{\nu}_e$ from top to bottom), and photons
(long dashed lines). 
Data are from Stanev\protect\cite{Stanev};
large crosses at EeV energies
are an estimate of the proton contribution to the total intensity
based on Fly's Eye observations.
(From Protheroe and Johnson\protect\cite{ProtheroeJohnson95}).
\label{fig:rachen_model}}
\end{figure}

Rachen and Biermann\cite{RachenBiermann93} have demonstrated that cosmic ray 
acceleration in Fanaroff-Riley Class II
radio galaxies can fit the observed 
spectral shape and the normalization at 10 -- 100 EeV
to within a factor of less than 10. 
The predicted spectrum below this energy also fits the proton
spectrum inferred from Fly's Eye data\cite{Rac93a}.
Protheroe and Johnson\cite{ProtheroeJohnson95} have repeated Rachen and Biermann's
calculation to calculate the flux of diffuse neutrinos and gamma rays which
would accompany the UHE cosmic rays, and
their result is shown in Fig.~\ref{fig:rachen_model}.

%%%%%%%%%%%%%%%%%%%%%%%%%%%%%%%%%%%%%%%%%%%%%%%%%%%%%%%%%%%%%%%%%%%%%%%%%%%%%%%%%%

\subsection{Observability of Ultra High Energy Gamma Rays}

Above 100 EeV the interaction properties of gamma rays 
in the terrestrial environment are very uncertain. 
Two effects may play a significant role: interaction with the
geomagnetic field, and the Landau-Pomeranchuk-Migdal (LPM) effect\cite{LPM1,LPM2}.

Energetic gamma rays entering the atmosphere will be subject to the
LPM effect (the suppression of electromagnetic cross-sections
 at high energy) which becomes very important at ultra-high energies. 
The radiation length changes as $(E/E_{\rm LPM})^{1/2}$, where 
$E_{\rm LPM} = 6.15 \times 10^4 \ell_{\rm cm}$ GeV, and 
$\ell_{\rm cm}$ is the standard Bethe-Heitler radiation length
in cm~\cite{StanevLPM82}.
Protheroe and Stanev\cite{ProtheroeStanev96} found
that average shower maximum will be reached 
 below sea level for energies $5 \times 10^{11}$ GeV, $8 \times 10^{11}$ GeV, 
and $1.3 \times 10^{12}$ GeV for gamma--rays entering the atmosphere
 at $\cos{\theta}$ = 1, 0.75, and 0.5 respectively. 
Such showers would
 be very difficult to reconstruct by experiments such as Fly's Eye
and at best would be assigned a lower energy.

Before entering the Earth's atmosphere  
gamma rays and electrons are likely to interact on the geomagnetic
 field (see Erber\cite{Erber68} for a review of the
 theoretical and experimental understanding of the interactions).
 In such a case the gamma rays propagating perpendicular to the geomagnetic
 field lines would cascade in the geomagnetic field, i.e. pair production
 followed by synchrotron radiation. 
The cascade process would degrade the gamma ray energies to some extent
(depending on pitch angle), and the atmospheric cascade would then
 be generated by a bunch of gamma-rays of lower energy. 
Aharonian et al.\cite{AharonianKanevskySahakian91} have
considered this possibility and conclude that this bunch would
appear as one air shower made up of the superposition of many air showers of 
lower energy, where the LPM effect is negligible; the air shower 
having the energy of the initial gamma-ray outside the geomagnetic field.
If this is the case, then gamma rays above 300 EeV
would be observable by Fly's Eye, etc.
There is however, some uncertainty as to whether pair production will take 
place in the geomagnetic field.
This depends on whether the geomagnetic field spatial
dimension is larger than the formation length of the electron
pair, i.e. the length required  to achieve a separation between the
two electrons that is greater than the classical radius of the electron
(see also Stanev and Vankov \cite{StanevVankov96}). 

\section{Exotic Origin Models}

We now discuss
the possibility that the highest energy cosmic rays are not single nucleons.
Obvious candidates are heavier nuclei (e.g. Fe),
$\gamma$-rays, and neutrinos. 
In general it is even more difficult to propagate
nuclei than protons, because of the additional photonuclear disintegration
which occurs\cite{Tka75,Pug76,Elb95}.
The possibility that the 300 EeV event is a $\gamma$-ray
has been discussed recently \cite{Hal94} 
and, although not completely ruled out,
the air shower development profile
seems inconsistent with a $\gamma$-ray primary.
Weakly interacting particles such as neutrinos will have no difficulty
in propagating over extragalactic distances, of course. 
This possibility has 
been considered, and generally discounted\cite{Hal94,Elb95},
mainly because of the relative unlikelihood of a neutrino interacting
in the atmosphere, and the necessarily great increase in the 
luminosity required of cosmic sources.
Magnetic monopoles accelerated by magnetic fields in our Galaxy
have also been suggested\cite{Kep95} and can not be ruled out 
as the highest energy events until the expected air shower development
of a monopole-induced shower is worked out.

\subsection{Topological Defect Origin}

  One exciting possible explanation of the highest energy
 cosmic rays is the topological defect (TD) 
scenario\cite{AharonianBhatSchramm92,BhatHillSchramm92,GillKibble94},
 where the
 observed cosmic rays are a result of top-down cascading,
 from somewhat below GUT scale energy, $\sim 10^{16}$ GeV, 
to $10^{11}$ GeV and lower energies.
Generally, these models put out much of the energy in a very flat
spectrum of photons and electrons extending up to the mass of 
``X--particles'' emitted which may be lower than $10^{16}$ GeV, 
depending on the theory.
Approximating this spectrum by a monoenergetic injection of photons of energy
$10^{15}$ GeV, Protheroe \& Johnson\cite{ProtheroeJohnson95} showed
that spectra from single TD sources can not explain the
$3 \times 10^{11}$ GeV events.

  The main problem with topological defect models is the wide range
 of model parameters in which this scenario could, in principle, be
 applied. This is very different from astrophysical
 scenarios\cite{RachenBiermann93} where nearly all of the
 properties of the astrophysical objects are restricted from
 observations (luminosities at different wavelengths, magnetic
 field, etc.).
 Parameters of TD scenarios include:
 mass of the X--particle (maximum injection energy for X--particle
 decay products), energy spectra and final state composition
 of the decay products, and cosmological evolution of the topological
 defect injection rate\cite{Sigl95,Lee96}.

 To compare the TD decay products with the cosmic ray data
 one has to study the change
 of the spectra during propagation\cite{ProtheroeJohnson95,Lee96}.
This problem is more severe than for the case of ``bottom-up'' scenarios
because most of the energy from X--particle decay,
and subsequent decays, emerges in electrons, photons and neutrinos,
with only about 3\% in nucleons.
The electrons and photons initiate electromagnetic cascades in the
extragalactic radiation fields and magnetic field, resulting in
a complicated spectrum of electrons and photons which is very
sensitive to the radiation and magnetic environment.
For example, recently the HEGRA group\cite{Karle95} have placed an
upper limit on the ratio of gamma-rays to cosmic rays of $\sim 10^{-2}$
at $10^5$ GeV, and based on a TD model  
calculation\cite{AharonianBhatSchramm92} which neglected the IR background
and gave a higher ratio argued that TD models were ruled out.
However, inclusion of the 
IR reduces the $10^5$ GeV gamma-ray intensity to below the HEGRA limit.

Unfortunately, many of the relevant
 parameters of extragalactic space (average magnetic field,
 strength of the radio and IR/optical backgrounds)
 are not well known.
Protheroe and Johnson\cite{ProtheroeJohnsonTaup95} have considered 
one set of parameters ($m_Xc^2=10^{15}$ GeV, constant injection per
co-moving volume, $B=10^{-9}$ gauss) and rule out TD as the origin
of the $3 \times 10^{11}$ GeV events.
More recently, Lee\cite{Lee96} and Sigl, Lee and Coppi\cite{SiglLeeCoppi96},
adopting a probably unphysical lower X--particle mass $m_Xc^2=10^{14}$ GeV 
(the X--particle mass is expected to be near the unification mass which is 
$10^{16.0 \pm 0.3}$ GeV \cite{Amaldi91}), 
and a lower magnetic field claim the TD scenario is not ruled out.
However, Protheroe and Stanev\cite{ProtheroeStanev96} argue that
these X--particles masses are ruled out as well.
The details are not simple, and include considering whether or not
UHE gamma rays in the cascade are observable or not by air shower arrays,
but in either case TD models appear to be ruled out {\it as the origin
of the 300 EeV events}.

%%%%%%%%%%%%%%%%%%%%%%%%%%%%%%%%%%%%%%%%%%%%%%%%%%%%%%%%%%%%%%%%%%%%%%%%%%%%%%%%%%

\section{Conclusion}

While the origin of the highest energy cosmic rays remains uncertain, 
there appears to be no necessity to invoke exotic models.
Shock acceleration, which is believed to be responsible for the
cosmic rays up to at least 100 GeV, is a well-understood mechanism 
and there is evidence of shock acceleration taking place in radio galaxies
\cite{BS87}, and the conditions there are
favourable for acceleration to at least 300 EeV.
Such a model, in which cosmic rays are accelerated in Fanaroff-Riley Class II
radio galaxies, can readily account for the flat component of cosmic rays
which dominates the spectrum above $\sim 10$ EeV \cite{Rac93a}.
Indeed, one of the brightest FR II galaxies, 3C 134, is a candidate 
source for the 300 EeV Fly's Eye event (P.L. Biermann, personal communication).

Whatever the source of the highest energy cosmic rays, because of their
interactions with the radiation and magnetic fields in the universe, the
cosmic rays reaching Earth will have spectra, composition and arrival
directions affected by propagation.
Many of the important parameters, e.g. extragalactic magnetic fields
and radiation fields, are uncertain, and more work is needed to obtain
better estimates of these and thereby help unravel the puzzle.
When better information is available, and good statistics on the arrival
directions, energy spectra and composition, as well as the intensity
of the diffuse background of very high energy gamma rays (partly produced
in cascades initiated by cosmic ray interactions), we will be better
placed to understand the origin of the highest energy particles occurring
in nature.
The Auger Project, an international collaboration to build two UHE
cosmic ray detectors, one in the United States and one in Argentina,
each having a collecting area in excess of 1000 km$^2$
(see the lecture by J.M. Matthews in this volume), will go a long
way to help answer these questions.

\section*{Acknowledgments}
I thank Maurice Shapiro for allowing me to participate as a student
in the 1st School of Cosmic Ray Astrophysics held in Erice, 
for inviting me to lecture at the 10th School, and together with John Wefel
for organizing an outstanding meeting.
I thank  Wlodek Bednarek, David Bird, and Qinghuan Luo for reading the 
manuscript.
My research is supported by a grant from the Australian Research Council.

\end{document}